\documentclass[aps,prl,twocolumn,superscriptaddress,groupedaddress]{revtex4} 
\usepackage {graphicx}  
\usepackage {dcolumn}   
\usepackage {bm}        
\usepackage {amssymb}   
\usepackage {color}
\usepackage {amsmath}
\usepackage {amsfonts}
\usepackage {amsthm}
\usepackage {mathrsfs}
\usepackage {natbib}
\usepackage {latexsym}
\usepackage {dsfont}
\usepackage {txfonts}
\usepackage {rotating}
\usepackage {multirow}
\usepackage {hhline}
\usepackage {hyperref}
\usepackage {bm}
\usepackage {appendix}
\usepackage {url}
\usepackage {acronym}
\usepackage {enumitem}
\usepackage {aas_macros}
\hypersetup{
  colorlinks=true,        
  linkcolor=black,         
  citecolor=cyan,         
}

\newcommand{\allpar}{\bm{\theta}}
\newcommand{\EMdat}{\bm{S}}
\newcommand{\GWdat}{\bm{D}}
\newcommand{\com}{\bm{\gamma}}
\newcommand{\GWnon}{\bm{\omega}}

\newcommand{\EMnon}{\bm{\phi}}

\begin{document}
\title{Probing intrinsic properties of short gamma-ray bursts with
gravitational waves}

\author{Xilong Fan} 
\email[Xilong Fan:]{Xilong.Fan@Glasgow.ac.uk}
 \affiliation{School of Physics and Electronics Information, Hubei University of Education, 430205 Wuhan, China}
\affiliation{SUPA, School of Physics and Astronomy, University of
  Glasgow, Glasgow G12 8QQ, United Kingdom}
  \author{Christopher Messenger}
\author{Ik Siong Heng}
\affiliation{SUPA, School of Physics and Astronomy, University of
  Glasgow, Glasgow G12 8QQ, United Kingdom}


\begin{abstract}
%
%
Progenitors of short gamma-ray bursts are thought to be neutron stars
coalescing with their companion black hole or neutron star, which are one of the main
gravitational wave sources. We have devised a
Bayesian framework for combining gamma-ray burst and gravitational wave
information that allows us to probe short gamma-ray burst luminosities. We show
that combined short gamma-ray burst and gravitational wave observations not
only improve progenitor distance and inclination angle estimates, they also
allow the isotropic luminosities of short gamma-ray bursts to be determined
without the need for host galaxy or light-curve information. We characterise
our approach by simulating 1000 joint short gamma-ray burst and gravitational
wave detections by Advanced LIGO and Advanced Virgo. We show that
${\sim}90\%$ of the simulations have uncertainties on short gamma-ray burst
isotropic luminosity estimates that are within a factor of 2 of the ideal
scenario, where the distance is known exactly. 
Therefore, isotropic luminosities can be confidently determined for 
short gamma-ray bursts observed jointly with gravitational wave detected by Advanced
LIGO and Advanced Virgo. Planned enhancements to Advanced LIGO will extend its
range and likely produce several joint detections of short gamma-ray bursts and
gravitational waves. Third-generation gravitational wave
detectors will allow for isotropic luminosity estimates for the majority of the
short gamma-ray burst population within a redshift of $z{\sim}1$.
\end{abstract}


\maketitle

\acrodef{GW}[GW]{gravitational-wave}
\acrodef{BNS}[BNS]{binary neutron star}
\acrodef{CBC}[CBC]{compact binary coalescence}
\acrodef{HMNS}[HMNS]{hypermassive neutron star}
\acrodef{SGRB}[SGRB]{short-duration gamma-ray burst}
\acrodef{LGRB}[LGRB]{long-duration gamma-ray burst}
\acrodef{sGRB}[sGRB]{short gamma-ray burst}
\acrodef{ET}[ET]{Einstein Telescope}
\acrodef{NS}[NS]{neutron star}
\acrodef{EM}[EM]{electromagnetic}
\acrodef{SNR}[SNR]{signal-to-noise ratio}
\acrodef{PDF}[PDF]{probability distribution function}
\acrodef{EOS}[EOS]{equation of state}
\acrodef{GWGC}[GWGC]{Gravitational Wave Galaxy Catalogue}
\acrodef{UNGC}[UNGC]{Updated Nearby Galaxy Catalog}
\acrodef{MMPF}[MMPF]{multi-messenger prior function}
\acrodef{OKDE}[OKDE]{online kernel density estimation}

\emph{Introduction}---
%
%
The most likely candidate for the progenitor of an \ac{sGRB} event is the merger
of a binary neutron
star~\citep{1989Natur.340..126E,2014ARA&A..52...43B}.  In
this scenario such an event will be accompanied by the emission of a
characteristic \ac{GW} signal detectable by the Advanced LIGO-Virgo \ac{GW}
interferometers as they approach design sensitivity~\cite{Aasi:2013wya}. A
binary neutron star observation would follow the recent historic first
detections of \acp{GW} from  binary black hole
mergers~\cite{2016PhRvL.116f1102A,2016PhRvX...6d1015A,2017PhRvL.118v1101A}. These merger events,
whilst not consisting of neutron star components, were actively followed-up via
multiple \ac{EM} channels~\cite{2016ApJ...826L..13A} including  gamma-ray
telescopes~\cite{2016ApJ...823L...2A,2016A&A...593L..10B,2016ApJ...826L...6C}.

%
Due to the moderately well constrained properties of \acp{sGRB}, they are
observed for only a narrow range of source orientations relative to the
observers line of sight. Since \acp{GW} from \ac{CBC} events are emitted
broadly isotropically the expected rate of joint \ac{sGRB}--\ac{GW} detections
is relatively low. This event rate for coincident \ac{sGRB}--\ac{GW}
events has been discussed in
\cite{1993ApJ...417L..17K,1999PhRvD..60l1101F,2013PhRvL.111r1101C,2013PhRvD..87l3004K,
2013PhRvD..87f4033D,2014MNRAS.445.3575C,2015ApJ...809...53C,2016MNRAS.459..121C,2016ApJ...827...75L,2016JCAP...11..056P,2017ApJ...840...88C}.

%
Beyond rate estimates, there have also been a number of studies, including work
presented in this paper, exploring how joint \ac{sGRB}--\ac{GW} detections can be
used to enhance understanding of the underlying physical system. It has been shown
by~\citep{2007PhRvD..75b4016S} that a three-dimensionally localized (sky
direction and distance) \ac{sGRB} in conjunction with the detection of the
\ac{GW} signal can improve the estimation of the inclination angle of the
\ac{CBC}. 
Also, although not truly a joint
analysis,~\cite{2014PhRvD..90b4060A} discusses how a multi-detector \ac{GW}
network alone can remove observational degeneracies in the inclination angle
measurement allowing us to better understand off-axis \ac{sGRB} events.

%

%
In this paper, we focus on what can we learn from a single \ac{sGRB}--\ac{GW}
detection. We model the system as being described by a set of parameters common
to both the \ac{sGRB} and \ac{GW} with additional parameters associated with
each phenomenon exclusively. Due to strong correlations between parameters, the
combination of information from each observation channel allows improvement in
common \emph{and} exclusive parameters ~\cite{2014ApJ...795...43F}. 
We specifically focus on the
improved estimation of the \ac{GW} inclination angle and the \ac{sGRB}
luminosity function in the likely scenario that there is a lack of an \ac{sGRB}
afterglow observation.

\emph{The statistical framework}---
%
%
We use the framework set out in~\cite{2014ApJ...795...43F} to set up our
analysis of joint \ac{EM}--\ac{GW} observations. Under this framework, for a
given set of \ac{sGRB} data, $\EMdat$, and \ac{GW} data, $\GWdat$, we divide
our observation parameters into three sets. The set of parameters common to
both sets of observations is denoted by $\com$, which for joint \ac{sGRB}--\ac{GW} 
observations are Right Ascension $\alpha$, declination $\delta$ and distance $d$. 
The parameters that are distinct to either only the \ac{EM} or \ac{GW} observations 
are denoted by $\EMnon$ and $\GWnon$ respectively. 

Our initial aim is to compute the posterior
distribution on all parameters $\allpar = (\com,\EMnon,\GWnon)$ conditional on
both datasets $\EMdat$, $\GWdat$ and any other implicit model assumptions
denoted by $I$. We start by using Bayes theorem to express the joint
distribution on the complete parameters set as
%
\begin{equation}\label{eq:joint_pos1} 
p(\allpar | \EMdat,\GWdat,I)=\frac{p(\com,\GWnon,\EMnon |I) p(\GWdat |
\com,\GWnon,I)p(\EMdat |\com,\EMnon,I)}{p(\EMdat,\GWdat | I)}.
\end{equation}
%
 In~\cite{2014ApJ...795...43F} a model
dependency was included that allowed the \ac{EM} parameters to govern the
probability that the \ac{GW} and \ac{EM} events originated from the same
source. In this analysis, we ignore this complexity and assume that all
\ac{sGRB} events have been uniquely associated with a \ac{GW} event.

%
The exclusive
unknown \ac{sGRB} parameters $\EMnon$ could contain many elements including the
time of arrival of the \ac{sGRB} in the detector frame, the duration of the
burst, and spectral parameters for example. For simplicity, we include only the
jet half opening angle $\theta_{\rm{jet}}$ and the isotropic \ac{sGRB}
luminosity $L_{\rm iso}$. Similarly, we have choices regarding the \ac{EM} data itself
and in this analysis we assume that our relevant measurement information is
contained within the measured peak flux $f_{\gamma}$. Hence, our \ac{EM} likelihood
becomes
\begin{equation}\label{eq:f_like_flux}
p(\EMdat|\com,\EMnon,I)=\frac{1}{\sigma_{f_\gamma}\sqrt{2\pi}}\exp
\left(-\frac{\left(f_{\gamma}- f_{\rm th}  \right)^2}{2\sigma_{f_\gamma}^2}
\right )
\end{equation}
where $f_{\text{th}}(d,L_{\rm iso},\theta_{\rm jet})$ is the expected value of the peak flux 

\begin{equation} \label{eq:f_th}
f_{\text{th}}(d,L_\mathrm{iso},\theta_{\rm jet})=\frac{L_{\rm iso}}{4\pi\,d^2}.
\end{equation}
%
%
%

%
The \ac{GW} likelihood is obtained from the standard parameter estimation
output of Bayesian Monte-Carlo sampling algorithms. The input \ac{GW} data
consists of time series of strain data output from multiple detectors within a
\ac{GW} interferometer network. The parameter estimation information is
available as a finite set of discrete samples drawn from the posterior
$p(\com,\GWnon|\GWdat,I)$. We note that in general this posterior therefore
already contains an assumed prior on the common and exclusive \ac{GW}
parameters which must be accounted for when converting back to the \ac{GW}
likelihood. The specific parameters included within the exclusive \ac{GW}
parameter set consist only of the neutron star masses $m_{1},m_{2}$, the
\ac{GW} polarisation angle $\psi$, and the binary inclination angle $\iota$.
Note that the inclination angle is linked to the jet half-opening angle since 
\acp{sGRB} are thought to be emitted perpendicular to the binary orbital plane 
(ie $\iota= 0$). 
Though these two quantities are related, they describe different physical
properties of the binary merger and are inferred from separate measurements. 
Thus, we do not listed them as one of the common parameters.

%
We assume all parameter priors to be independent with the exception of the jet
half opening angle and the binary inclination angle. They can therefore be
written as
\begin{equation}
\label{eq:priors}
p(\allpar) = p(m_{1}|I)p(m_{2}|I)p(\psi|I)p(d|I)p(L|I)p(\theta_{\rm jet},\iota|I)
\end{equation}
The correlation between the $\theta_{\rm jet}$ and $\iota$ priors is due to the
fact that we assume that our sources have been jointly detected and therefore
the jet half opening angle must therefore be greater than the inclination angle. We also
adopt the belief that the prior for the isotropic luminosity  follows a
power law distribution \cite{2004ApJ...601..371L} such that  
\begin{equation} \label{eq:l_prior} 
p(L_{\rm iso}|I) =\frac{0.4}{L_{\rm min}}
\left(\frac{L_{\rm iso}}{L_{\rm min}}\right)^{-1.4}.  
\end{equation}
with a lower cut-off luminosity $L_{\rm min}=10^{49}$erg s$^{-1}$ .

\emph{Simulation}---
%
%
To characterise our method, we generated \ac{GW} signals from \ac{BNS}
coalescences using
\emph{lalapps\_inspinj}\footnote{https://www.lsc-group.phys.uwm.edu/daswg/projects/lalsuite.html}.
The signal parameters are drawn from the prior defined in Eq.~\ref{eq:priors}
where masses are uniformly distributed on the range $(1.3,1.5)\,M_{\odot}$, the sky
position is uniform on the sky, $\psi$ is uniform on the range $(0,2\pi)$ and
distance is selected uniformly in volume out to a maximum distance of 460~Mpc,
equal to the horizon distance of the Advanced LIGO-Virgo network. At such
distances, the effect of cosmological redshift is minimal and we do not include
such effects in our simulations.  The joint distribution on $\theta_{\rm jet}$
and $\cos\iota$ is uniform under the constraints that $\cos\theta_{\rm
jet}<\cos\iota$ and $5^{\circ}<\theta_{\rm jet}<30^{\circ}$ with values
generated using rejection sampling. In total 1000 signals are generated and
added to simulated noise from a three-detector network consisting of the two
Advanced LIGO detectors (Hanford and Livingston) and Advanced Virgo at design
sensitivity~\citep[see more details in][]{2014ApJ...795...43F}.   

%
In order to construct an estimate of the \ac{GW} likelihood, a kernel density
estimation procedure\footnote{$\rm
http://www.vicos.si/Research/Multivariate\_Online\_Kernel\_Density\_Estimation$}
is used to compute the \ac{GW} likelihood term $p(\GWdat|\com,\GWnon,I)$ at any
location in the $(\com,\GWnon)$ parameter space. The discrete samples used as
input are those generated from the posterior distribution on the \ac{GW}
parameters obtained using
\emph{lalinference}~\citep{2013PhRvD..88f2001A,Veitch2010PhRvD..81f2003V}.
Since we require the likelihood, the prior distributions used in the generation
of the posterior must be removed. Since the $\cos\iota$ prior is uniform the
likelihood and posterior are directly proportional and no change is necessary.
The distance prior used to generate the posterior samples is uniform in volume
and hence $\propto d^2$. This is the same as that assumed for our general
analysis as defined in Eq.~\ref{eq:priors} and so the \ac{GW} posterior
represents the combined terms $p(\GWdat|\GWnon,\com,I)p(d|I)$. 

%
The \ac{EM} term $p(\EMdat|\com,\EMnon,I)$ in the numerator of 
Eq.~\ref{eq:joint_pos1} is computed under the assumption that the \ac{sGRB}
skymap is both consistent with, and significantly more constraining than the
\ac{GW} skymap. This allows us to treat the sky position as known. As stated
above, the jet half opening angle is drawn jointly with the binary inclination angle
consistent with the prior. The \ac{sGRB} luminosity value is drawn
from a power law distribution according to Eq.~\ref{eq:l_prior}. The measured flux $f_{\gamma}$
value  is then drawn from a Gaussian distribution (consistent with our assumed
likelihood, Eq.~\ref{eq:f_like_flux}) with  mean $f_{\rm th}$ given by
Eq.~\ref{eq:f_th} and a standard deviation equal to 30\% of the mean.

%
The final joint posterior distribution on all signal parameters
$p(\allpar|\EMdat,\GWdat,I)$ is obtained by computing the product of the
estimated \ac{GW} term and the analytic \ac{EM} term multiplied by the priors on
the all parameters according to Eq~\ref{eq:joint_pos1}. From this we can
compute marginalised posterior distributions on any parameter, for example, the 
isotropic luminosity posterior for a joint \ac{sGRB}-\ac{GW} detection is given by
\begin{equation}
\label{eq:Ljoint_post}
p(L_{\rm iso}|\EMdat,\GWdat,I) = \int d\allpar^{(\neq L_{\rm iso})}p(\allpar|\EMdat,\GWdat,I).
\end{equation}
where we have integrated over all parameters excluding the luminosity
(indicated by the superscript $\neq L_{\rm iso}$ on the $\allpar$ parameter set).
Such a distribution represents the combined inference power of both the
\ac{sGRB} and \ac{GW} observations.

\emph{Results}---
The improvements in the accuracy of sky location and trigger time via a joint
GW-SGRB observation have already been discussed in~\cite{2015ApJS..217....8B}.
Here, we focus on the inference of distance, inclination angle and \ac{sGRB} isotropic luminosity.

\begin{figure} 
\begin{center}
\includegraphics[width=\columnwidth]{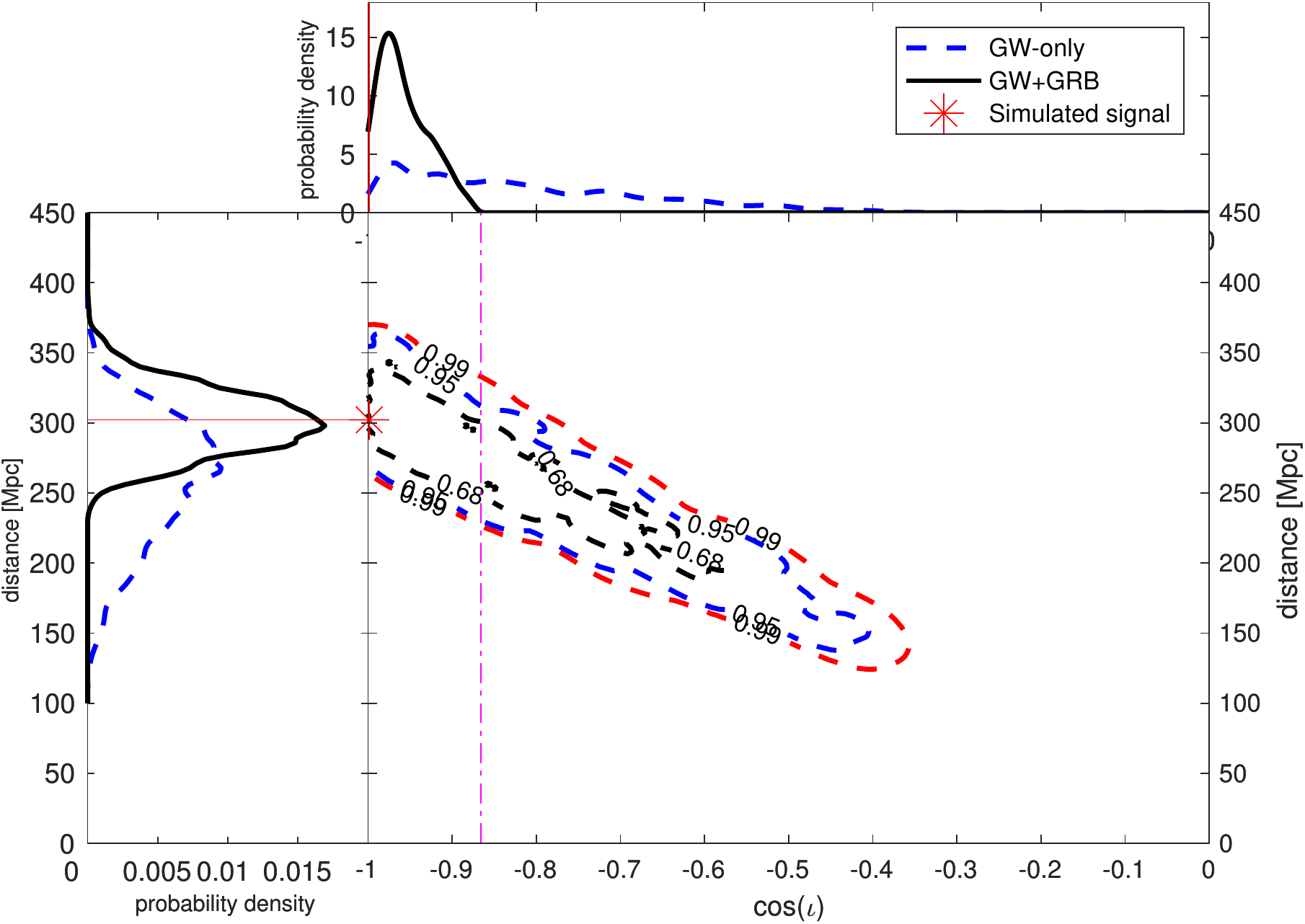}
\caption{Joint and marginalised posterior distributions on the distance and
cosine inclination for a \ac{GW} analysis and a joint \ac{sGRB}-\ac{GW} analysis.
The example shown here is for a source at $d=302$ Mpc, $\cos\iota= -0.99$,
$\theta_{\rm jet}=14.44^{\circ}$, $L_{\rm iso}=5\times10^{49}$ erg. The distance and $\cos\iota$ values
are indicated on the figure by an asterisk. The corresponding \ac{GW}
signal-to-noise ratio is 15.88. Vertical lines in the marginalised posterior
plots indicate the true simulated values. The restrictions imposed by the \ac{sGRB} jet
half opening angle (vertical dash-dotted  lines) cause significant reduction in the distance and inclination
angle uncertainties.}\label{fig:d_cos}
\end{center} \end{figure}

%
We select one of the simulations as a case study for illustrating the effects
of the joint analysis (signal parameters given in Fig.~\ref{fig:d_cos}).
Contours representing the joint posterior on distance and $\cos\iota$ are
plotted in Fig.~\ref{fig:d_cos} together with their marginalised distributions.
The joint \ac{sGRB}-\ac{GW} analysis allows us to apply jet half opening angle
priors which constrain the inclination angle and consequently significantly
improves the distance estimate. The combined \ac{sGRB}-\ac{GW} posterior shown
in Fig.~\ref{fig:d_cos} are produced by applying
Eqs.~\ref{eq:f_like_flux}-\ref{eq:l_prior} and are not obtained by the direct
application of a threshold on the half opening angle posterior.  For this particular
case study, the 95$\%$ credible intervals for distance and $\cos\iota$ are
improved by factors of $\sim$ 2.5 and 8, respectively.

%
%

\begin{figure}
\begin{center} 
\includegraphics[width=\columnwidth]{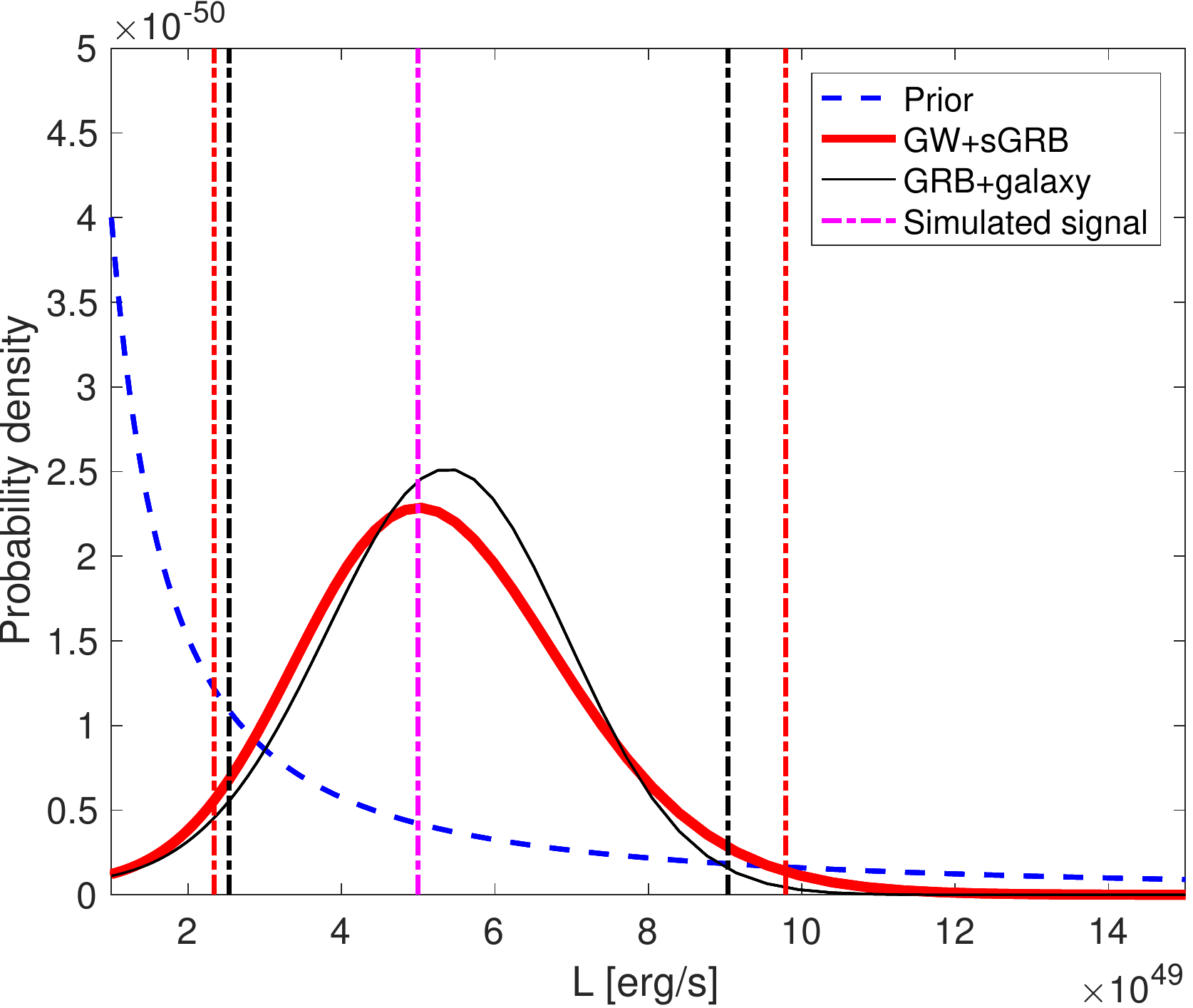}
\caption{The marginalized probability density on \ac{sGRB} isotropic luminosity for our
case study. The bold red curve shows the result for a joint \ac{sGRB}-\ac{GW}
observation and the thin blue curve shows the result for an \ac{sGRB} observation
with an identified  host galaxy and corresponding exactly known distance.  The
vertical  red and black dash-dotted lines  represent the the 95$\%$
credible regions for each case respectively.  The blue dashed curve shows the assumed
prior for both cases (given by Eq.~\ref{eq:l_prior}). The simulated value of
isotropic luminosity is indicated by the vertical magenta dash-dotted line.}\label{fig:l_est} 
\end{center}  
\end{figure}

In Fig.~\ref{fig:l_est} we show the marginalized probability density on the
\ac{sGRB} isotropic luminosity in our case study for 2 different scenarios.  The first is
the luminosity posterior assuming a joint \ac{sGRB}-\ac{GW} detection where we
have marginalised all parameters excluding the luminosity using
Eq.~\ref{eq:Ljoint_post}. The second curve is the posterior obtained using only
an \ac{sGRB} detection together with a correctly identified host galaxy at the
true distance $d'$.  In this case the distance is then assumed known and the
corresponding luminosity posterior is given by
\begin{equation} 
\label{eq:LGRB_post} 
p(L_{\rm iso}|\EMdat,d{=}d',I) \propto \iint
p(\EMdat|\EMnon,\com,I)p(\EMnon,\com|I)\delta(d-d')\,d\EMnon d\com.
\end{equation}
In this case study the 95$\%$ credible intervals show that the joint
\ac{sGRB}-\ac{GW} luminosity estimation is comparable with that of the 
\ac{sGRB}-host galaxy observation.

%
Currently, isotropic luminosity estimates for \acp{sGRB} rely on obtaining redshift
measurements of their assumed host galaxies. Only $\sim 30\%$ of all detected
\ac{sGRB} have an identified host galaxy \footnote{This percentage is also consistent with counting entries with redshift estimates at \url{https://swift.gsfc.nasa.gov/archive/grb_table/}.} (\cite{2012ApJ...746...48M}), while all \acp{sGRB}
observed in conjunction with \ac{GW} counterparts will have a distance estimate
directly from the \ac{GW} observation. For \acp{sGRB} with identified host
galaxies, the flux measurement uncertainty contributes to the spread in the isotropic
luminosity posterior. For an \ac{sGRB}-\ac{GW} observation, all additional
posterior width is due to the uncertainty remaining in the distance after the
degeneracy between inclination angle and distance has been constrained by the joint
\ac{sGRB}-\ac{GW} detection.

%
We examine the effectiveness of our proposed joint \ac{sGRB}-\ac{GW} analysis
by examining the posterior credible intervals for the inferred
source distance and \ac{sGRB} isotropic luminosity using 1000 simulated signals as
described previously.

\begin{figure*}
\begin{center}
\includegraphics[width=0.45\textwidth]{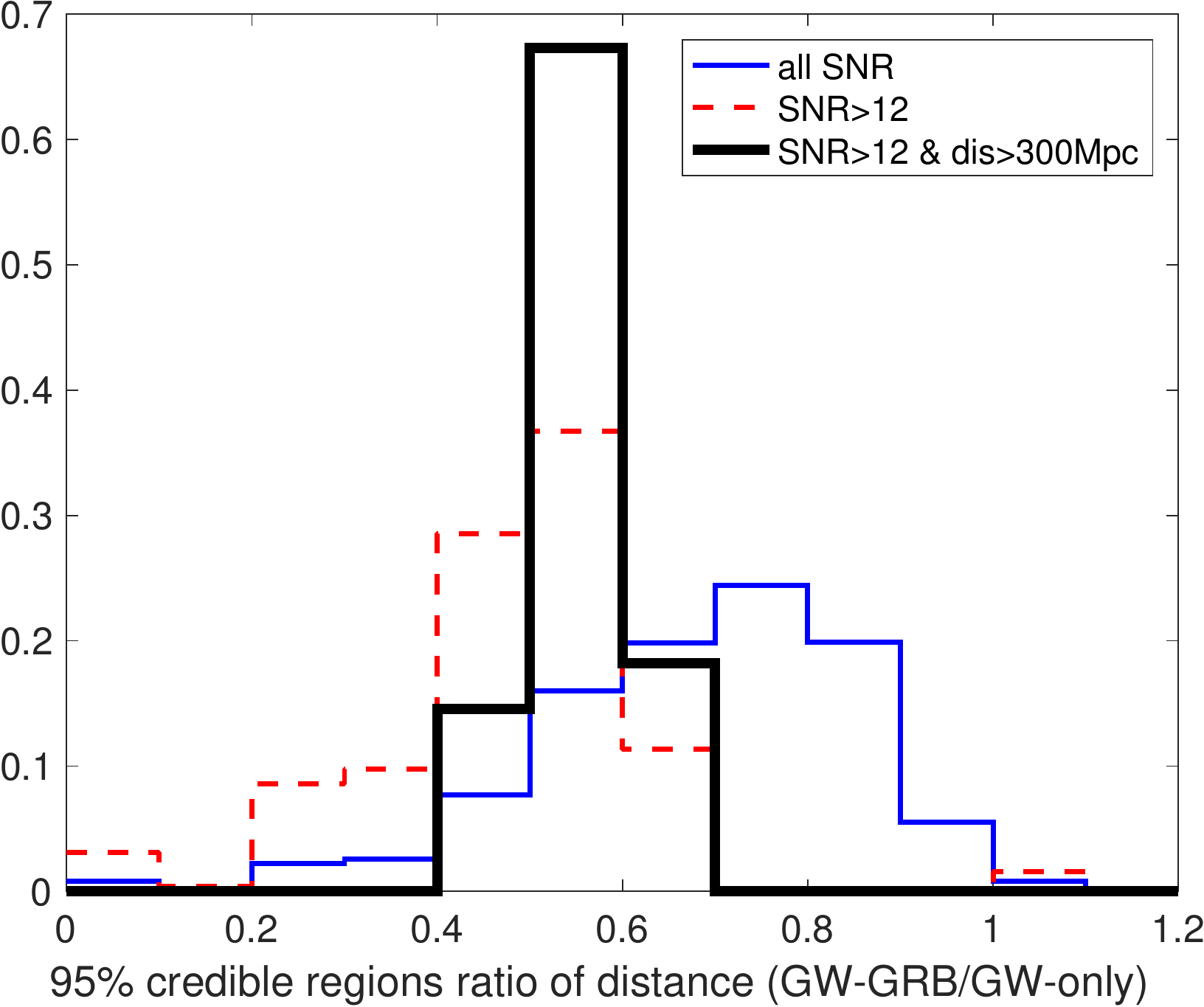}
\includegraphics[width=0.45\textwidth]{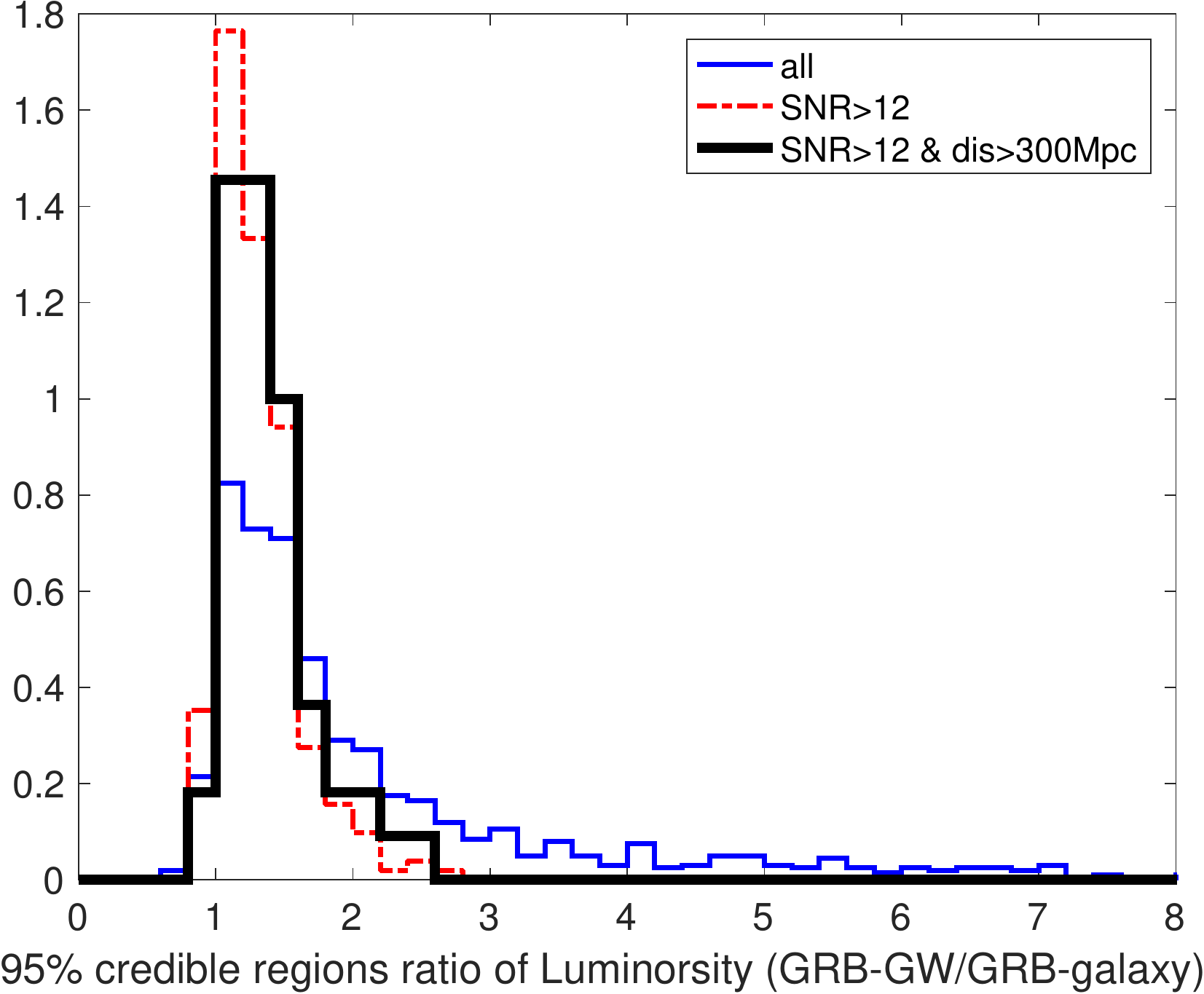}
\caption{(left) Histogram of the $95\%$ credible interval ratios for distance.
We take the ratios of the credible intervals between the joint
\ac{sGRB}-\ac{GW} observations and the \ac{GW} only scenarios.  (right)
Histogram of the $95\%$ credible interval ratios for isotropic luminosity. We take the
ratios of the credible intervals between the joint \ac{sGRB}-\ac{GW}
observations and the \ac{sGRB}-host galaxy scenarios. In both plots we consider
3 subsets of the results: all results (1000 samples, blue), \ac{SNR}~${>}12$
(256 samples, red) and \ac{SNR}~${>}12$ \& distance~${>}300$~Mpc (55 samples,
black). All cases are normalized according to their respective sample
sizes. Note that the majority of the credible regions on isotropic luminosity obtained with the 
\ac{sGRB}-\ac{GW} analysis are similar to those obtained using host galaxy identification,
 with credible region ratios between $\sim 1$ and $2$.}\label{fig:hist_dl_est} 
\end{center} 
\end{figure*}

%
In Fig.~\ref{fig:hist_dl_est}, we compare the $95\%$ credible intervals on the
distance posterior distributions obtained using joint \ac{sGRB}-\ac{GW}
observations with those obtained for \ac{GW} observations alone. 
Observing a \ac{sGRB} in conjunction
with a \ac{GW} provides an additional constraints on the inclination angle
which reduces the uncertainty in the source distance estimation. The $95\%$
credible regions can be as much as $\sim$ 10 times smaller for joint
\ac{sGRB}-\ac{GW} analyses. In general, however,   the median ratio  is $0.65$. The
fiducial detection threshold for ground based detectors is \ac{SNR} ${\sim}12$ and if
we restrict our analysis to only consider signals above this value then we find that the
distance uncertainty reduction has a narrower range and shifts to lower values
with a median ratio of $0.45$. The fact that it is most likely that a
joint \ac{sGRB}-\ac{GW} observation will occur close to the horizon distance of
the \ac{GW} detector network is already accounted for in the distance prior.
However, if we further restrict the analysis to detectable but distant signals
such that the \ac{SNR}~$>12$ and the distances are $>300$ Mpc then the
median ratio  becomes  $0.50$. 

%
We also examine the ability of the joint \ac{sGRB}-\ac{GW} analysis to infer
the \ac{sGRB} isotropic luminosity and compare it to luminosities inferred using distances
from identified \ac{sGRB} host galaxies. We have assumed that all 1000
simulated signals have identified host galaxies and the host galaxy distances
have been measured exactly. While this scenario is favorable for isotropic
luminosity estimates obtained via host galaxy identification, we note that the
host galaxy distance estimates can have a broad range of uncertainties (e.g.
photometric redshifts would have greater uncertainty than spectroscopic
redshifts, and host galaxies can be incorrectly identified) and can lead to
considerable uncertainties in the corresponding luminosity estimate. However,
to be conservative, we choose to compare scenarios that favour the \ac{sGRB}
host galaxy approach. With this in mind, in Fig.~\ref{fig:hist_dl_est}, we
compare the $95\%$ credible intervals for the isotropic luminosity posterior
distributions from the joint \ac{sGRB}-\ac{GW} analysis with that for source
luminosities obtained via perfect galaxy host identification and corresponding
error-free redshift measurements. We see that the credible intervals for the
joint \ac{sGRB}-\ac{GW} analysis are typically within a factor of two of that
obtained using the host galaxies. Given our optimistic assumptions regarding
host galaxy identification, the relative level of uncertainty in the luminosity
estimate achieved by the joint \ac{sGRB}-\ac{GW} analysis shows that we can
obtain a reliable isotropic luminosity estimate for most \ac{sGRB}-\ac{GW} observations
using our proposed method.

\emph{Discussion}---
%
%
We have proposed a general procedure for the parameter estimation of two
independent observations and applying it to reveal the nature of \acp{sGRB}
using joint \ac{sGRB}--\ac{GW} observations. We have quantified the level to
which the distance--inclination angle degeneracy inherent to \ac{GW}
observations can be broken with the inclusion of \ac{sGRB} data. In addition to
this, we have shown that through this joint analysis and the
distance--inclination angle degeneracy breaking we are able to estimate the
isotropic luminosity of the \ac{sGRB}. The precision with which we are able to
do this is comparable with the precision possible for \ac{sGRB} analyses
without \ac{GW} counterparts but with a well-defined host galaxy.

%

%
A unique feature of our approach is that it will provide isotropic luminosity
estimate for every joint \ac{sGRB}--\ac{GW} detection. This is in contrast to
the fact that only $\sim$30\% of \ac{sGRB} events have redshift estimates and
hence an isotropic luminosity estimate. In the near term, the Advanced ground based
\ac{GW} detectors are likely to detect only a handful of joint
events~\cite{2015ApJ...809...53C} and hence only make a small contribution to
the $O(100)$ that will be known via \ac{sGRB} observations on the same
timescale. However, 3rd generation \ac{GW}
detectors~\cite{2012CQGra..29l4013S,whitepaper} will be sensitive to \ac{BNS}
systems out to redshift $z\sim 1$ and will therefore jointly detect $\sim$ all
\ac{sGRB} events out to this distance.  Joint detections will therefore provide
$\sim$twice as many luminosity measurements as are likely with \ac{sGRB} events
alone (assuming no improvement in redshift estimation). 

%
By combining the information from \ac{EM} and \ac{GW} channels we have been
able to quantify the improvements possible in the merger distance estimation.
The spread in improvement ranges between a factor of 1--8 for all simulations
and those detectable with \ac{SNR}~${>}12$ showing improvements clustered around
factors of $\sim 2$.

%
One possible extension to this work is to embed this analysis inside a
hierarchical Bayesian model with the aim of inferring the \ac{sGRB} isotropic luminosity
distribution. In this work we have assumed a power-law prior distribution of
the luminosity prior and the corresponding power-law index. A hierarchical
procedure could be used to estimate this index and other parameters like the
lower luminosity cut-off. It could also be used to perform model selection
between, for example, single and broken power-law models. Additionally,
non-parametric techniques such as Gaussian process modelling could provide
estimates for the form of the isotropic luminosity distribution. 

%
The method we have introduced in this paper is not just applicable to joint
observations using \acp{sGRB} and \acp{GW}. The power of the joint inference we
describe can be applied to any multi-messenger observations (2 or more and not
necessarily including \ac{GW} observations) and will naturally exploit the
parameter correlations between common parameters. As shown here, such
correlations can lead to improved inference on other system parameters, and
could be applied to further astrophysical phenomena associated with \ac{GW}
events such as supernovae, kilonovae, or high energy neutrino observations.

\emph{Acknowledgements}---We would like to acknowledge valuable input from M.~Hendry, F.~Pannarale, D.~Holz and N.~Tanvir. The
authors also gratefully acknowledge the support of this research by the Royal
Society, the Scottish Funding Council, the Scottish Universities Physics
Alliance and and the Science and Technology Facilities Council of the United
Kingdom.  XF acknowledges financial support from National Natural Science
Foundation of China (grant No.~11673008,11633001).  CM and SH are supported by the
Science and Technology Research Council (grant No.~ST/~L000946/1).
%

%
\bibliography{gw_sgrb_paper.bbl}
%
%
%
%
%
\end{document}